\documentclass[aps,prl,amsmath,amssymb,superscriptaddress,lengthcheck]{revtex4-1}

\newcommand{\hftep}{$\textrm{Hf}_{2}\textrm{Te}_2\textrm{P}$}
\newcommand{\zrtep}{$\textrm{Zr}_{2}\textrm{Te}_2\textrm{P}$}
\usepackage{graphicx}
\usepackage{dcolumn}
\usepackage{bm}
\usepackage{relsize}
\usepackage{xcolor,color}
\usepackage{ulem}
\usepackage{hyperref}
\hypersetup{colorlinks=true,urlcolor=blue,citecolor=blue}
\begin{document}

\title{Topological surface states above the Fermi energy in \hftep.}

\author{T.\,J.\,Boyle}
\affiliation{Department of Physics, University of California, Davis, California 95616, USA}

\author{A.\,Rossi}
\affiliation{Department of Physics, University of California, Davis, California 95616, USA}
\affiliation{Advanced Light Source, Lawrence Berkeley National Lab, Berkeley, California 94720, USA}

\author{M.\,Walker}
\affiliation{Department of Physics, University of California, Davis, California 95616, USA}

\author{P.\,Carlson}
\affiliation{Department of Physics, University of California, Davis, California 95616, USA}

\author{M. K.\,Miller}
\affiliation{Department of Physics, University of California, Davis, California 95616, USA}

\author{J.\,Zhao}
\affiliation{Department of Physics, University of California, Davis, California 95616, USA}

\author{P.\,Klavins}
\affiliation{Department of Physics, University of California, Davis, California 95616, USA}

\author{C.\,Jozwiak}
\affiliation{Advanced Light Source, Lawrence Berkeley National Lab, Berkeley, California 94720, USA}

\author{A.\,Bostwick}
\affiliation{Advanced Light Source, Lawrence Berkeley National Lab, Berkeley, California 94720, USA}

\author{E.\,Rotenberg}
\affiliation{Advanced Light Source, Lawrence Berkeley National Lab, Berkeley, California 94720, USA}

\author{V.\,Taufour}
\affiliation{Department of Physics, University of California, Davis, California 95616, USA}

\author{I.\,Vishik}
\affiliation{Department of Physics, University of California, Davis, California 95616, USA}

\author{E.\,H.\,da Silva Neto}
\email[]{ehda@ucdavis.edu}
\affiliation{Department of Physics, University of California, Davis, California 95616, USA}

\begin{abstract}

We report a detailed experimental study of the band structure of the recently discovered topological material \hftep. Using the combination of scanning tunneling spectroscopy and angle-resolved photo-emission spectroscopy with surface K-doping, we probe the band structure of \hftep~with energy and momentum resolution above the Fermi level. Our experiments show the presence of multiple surface states with a linear Dirac-like dispersion, consistent with the predictions from previously reported band structure calculations. 
In particular, scanning tunneling spectroscopy measurements provide the first experimental evidence for the strong topological surface state predicted at $460$\,meV, which stems from the band inversion between Hf-\textit{d} and Te-\textit{p} orbitals. This band inversion comprised of more localized \textit{d}-states could result in a better surface-to-bulk conductance ratio relative to more traditional topological insulators.

\end{abstract}

\maketitle

Topological quantum phases have been discovered in a variety of materials such as topological insulators, Dirac and Weyl semimetals, and nodal line semimetals~\cite{Qi_2011,Hasan_2010,Ando_2013,Armitage_2018,Yang_2018}.
All of these topological phases result in non-trivial topological surface states (SSs), which 
could be used for future low-dissipation electronic or spintronic technologies
~\cite{Hasan_2010,Alicea_2012,Smejkal_2018}.
Although several materials that host one of the aforementioned topological phases have been discovered, quantum materials that exhibit several SSs at distinct regions of energy-momentum space are not common. 
Recently, band structure calculations and experiments indicate a rich phenomenology of Dirac-like electronic states in \zrtep~\cite{ZrTeP_2016} and in \hftep~\cite{Hosen_2018}, including multiple SSs.
In particular, for \hftep, the calculations find four Dirac-like SSs as follows, where the energies are relative to the Fermi level ($E_F$): three Dirac dispersions at the $\Gamma$ point near $0.46$\,eV (SS1), $0.17$\,eV (SS2), $-1.2$\,eV (SS3), and a Dirac node-arc along the $\Gamma$-M direction, centered at the M-point near $-0.9$\,eV (SS4). 
Similar features (SS1, SS2 and SS4) were also predicted in \zrtep. 
This unusual multitude of Dirac states in a single material stems from the topological character of various bulk bands originating from Te-\textit{p} and Hf-\textit{d} orbitals. Following Fu and Kane~\cite{Fu_2007}, the four topological $\mathbb{Z}_{2}$ invariants were computed from the calculated band structure. 
From this analysis, SS4 at $-0.9$\,eV was identified to  have a weak topological $\mathbb{Z}_{2}$ invariant, while SS1 at $0.46$\,eV yields a strong $\mathbb{Z}_{2}$ invariant. The topological nature of SS2 and SS3 could not be clearly resolved, with SS2 displaying a significant bulk-surface mixing \cite{Hosen_2018}. 

Among the SSs in \hftep, SS1 at $0.46$\,eV may be the one with the greatest potential toward future applications for several reasons. First, excluding SS2 due to its surface-bulk mixing, SS1 is the closest state to the Fermi level of \hftep.  
Second, it is the only SS centered within a clear direct gap in the bulk. Third, its topological nature originates from the inversion of Te-\textit{p} and Hf-\textit{d} orbitals, which could lead to a better surface to bulk conduction ratio when compared to more traditional topological insulators like Bi$_2$Te$_3$ or Bi$_2$Se$_3$, where only \textit{p}-orbitals are involved \cite{Yang2014, ZrTeP_2016, Hosen_2018}. This favourable surface conduction stems from the more localized nature of the Hf-\textit{d} (or Zr-\textit{d} in the case of \zrtep) states, leading to a higher effective mass for the bulk electrons when compared to Bi$_2$Te$_3$ or Bi$_2$Se$_3$, or even to the other SSs in \hftep~that do not involve Hf-\textit{d} orbitals. Additionally, this would render the interesting combination of topology and strong electron correlations, if $E_F$ is tuned to within the pseudogap.

Experimentally, the Dirac SSs in \zrtep~and \hftep~below the Fermi level, $E_F$, have been resolved through angle-resolved photo-emission spectroscopy (ARPES) measurements, showing a remarkable agreement to the band structure predictions~\cite{ZrTeP_2016,Hosen_2018}. However, since ARPES experiments are not able to probe the SS1 state above the Fermi level, this key topological state still lacks direct experimental evidence.

Here, we use two methods to experimentally access the states above $E_F$ and provide the direct evidence for SS1 in \hftep. First, through the deposition of K atoms we electron-dope the surface of the material, enabling ARPES to measure states up to approximately $160$\,meV above the original $E_F$. Second, we use scanning tunneling microscopy and spectroscopy (STM/S) to measure energy-resolved quasiparticle interference (QPI) patterns on the surface of \hftep, which allows us to probe the energy-momentum structure both above and below $E_F$~\cite{Hoffman_QPI_2002, Wang_Lee_QPI, Roushan_Nature_2009, Zeljkovic2014, Yazdani_review}. 
Using our combined ARPES and STM/S data, and their comparison to the reported surface band structure calculations, we resolve the SSs of \hftep~above $E_F$ and identify Dirac-like dispersive bands centered at $170$\,meV and $460$\,meV, where the SS1 and SS2 states have been predicted. 

\begin{figure*}
\includegraphics[width=180mm]{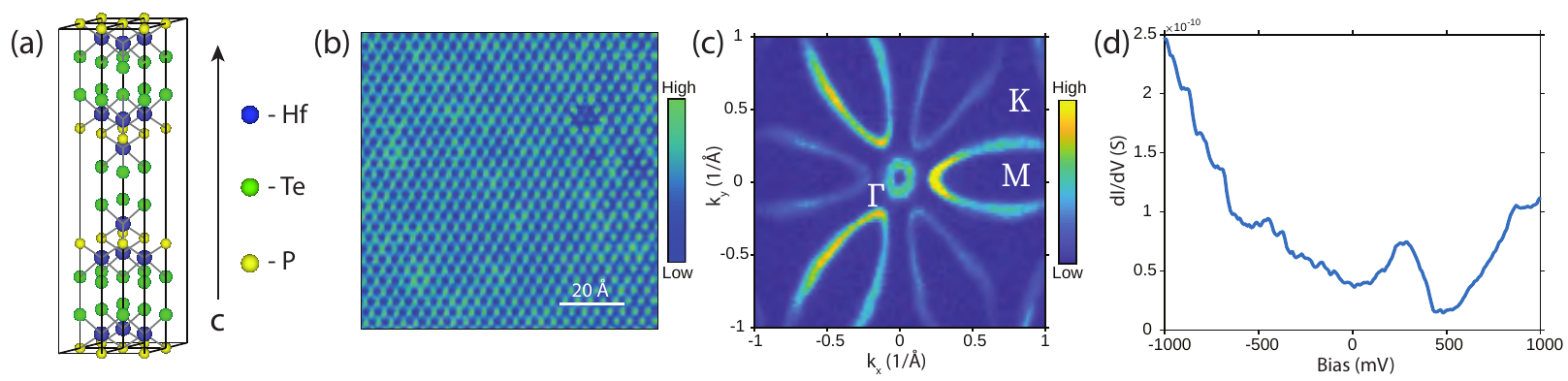}
\caption{\label{fig:1} (a) Crystal structure of \hftep. (b) STM topography taken at $-120$\,mV sample bias and $700$\,pA current setpoint. (c) Fermi surface of pristine sample measured by ARPES. (d) STS measured $dI/dV$ spectrum with current and bias set points of $450$\,pA and $-1.5$\,V.
}
\end{figure*}

Single crystals of \hftep\, were grown from vapor transport following the synthesis developed for the isostructural Zr$_2$Te$_2$P~\cite{Tschulik_2009, SM_}, also detailed in Ref.\,\cite{Chen_2016}. ARPES experiments were performed in ultra-high vacuum (UHV) at beamline 7.0.2 MAESTRO at the Advanced Light Source. The $\mu$-ARPES endstation allows for a spot size of $80$ $\mu$m x $80$ $\mu$m, which is crucial to scanning a homogeneous region in a cleaved area. After growth, the \hftep\, crystals were sealed in an inert environment, transferred into a glove box at the beamline and cleaved. From the glove box, they were transferred into the ARPES chamber without air exposure and cleaved a second time \textit{in situ}. The measurements were done at $77$\,K with a photon energy of $100$\,eV. The sample surface was doped with K using a commercial SAES alkali metal dispenser. STM/S measurements were done with a customized Unisoku USM-1300 instrument. The samples were cleaved \textit{in situ} in a UHV environment with pressures below $1\times10^{-9}$\,Torr. All STM/S measurements were performed at $4.2$\,K.

\begin{figure*}
\includegraphics[width=180mm]{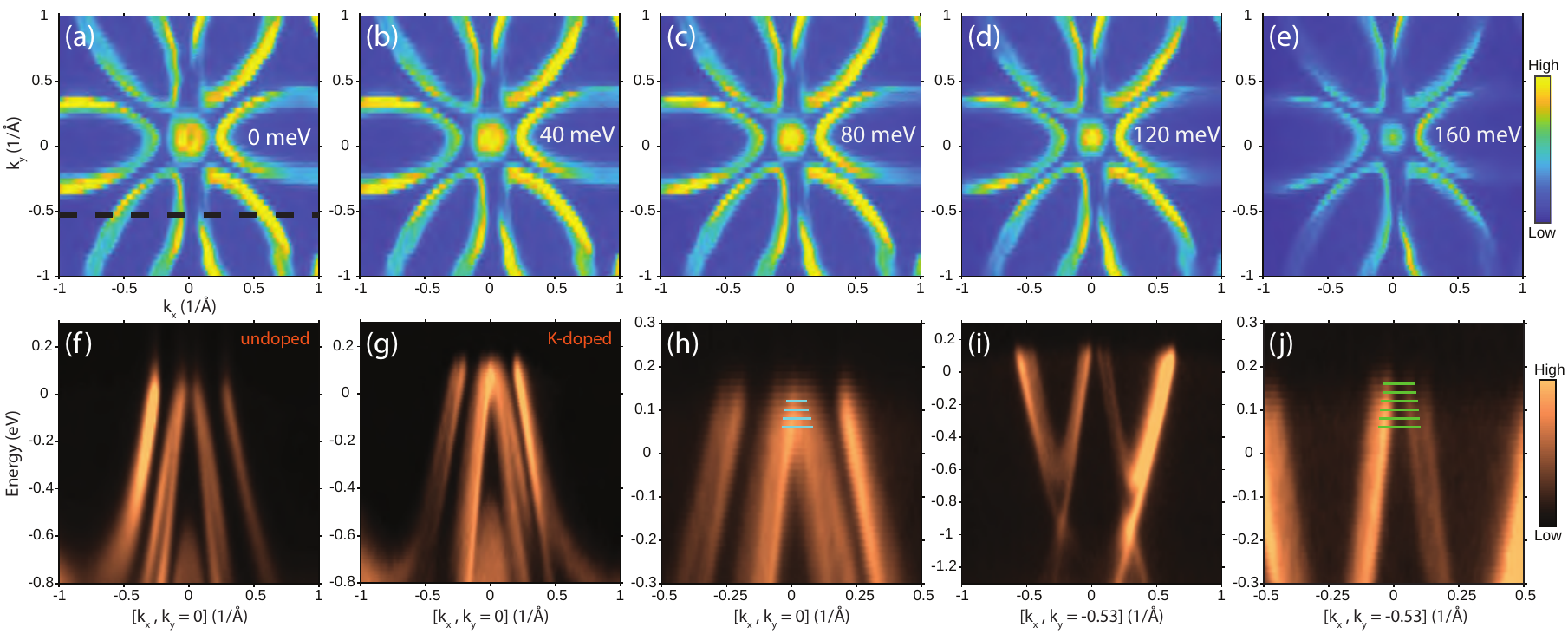}
\caption{\label{fig:2} (a-e) ARPES measured constant-energy contours of the K-doped sample for different positive energy values. (f) Dispersion curve of the undoped sample along $\Gamma$-M. (g) Dispersion curve of the doped sample along $\Gamma$-M. (h) Zoomed in version of (g) showing the Q-vectors (cyan lines) associated with the hole pocket centered at $\Gamma$ for four different energies. (i) Dispersion curve along the direction of the dotted line in (a). (j) Zoomed in version of (i) showing the Q-vectors (green lines) for five different energies.
}
\end{figure*}

\begin{figure*}
\includegraphics[width=180mm]{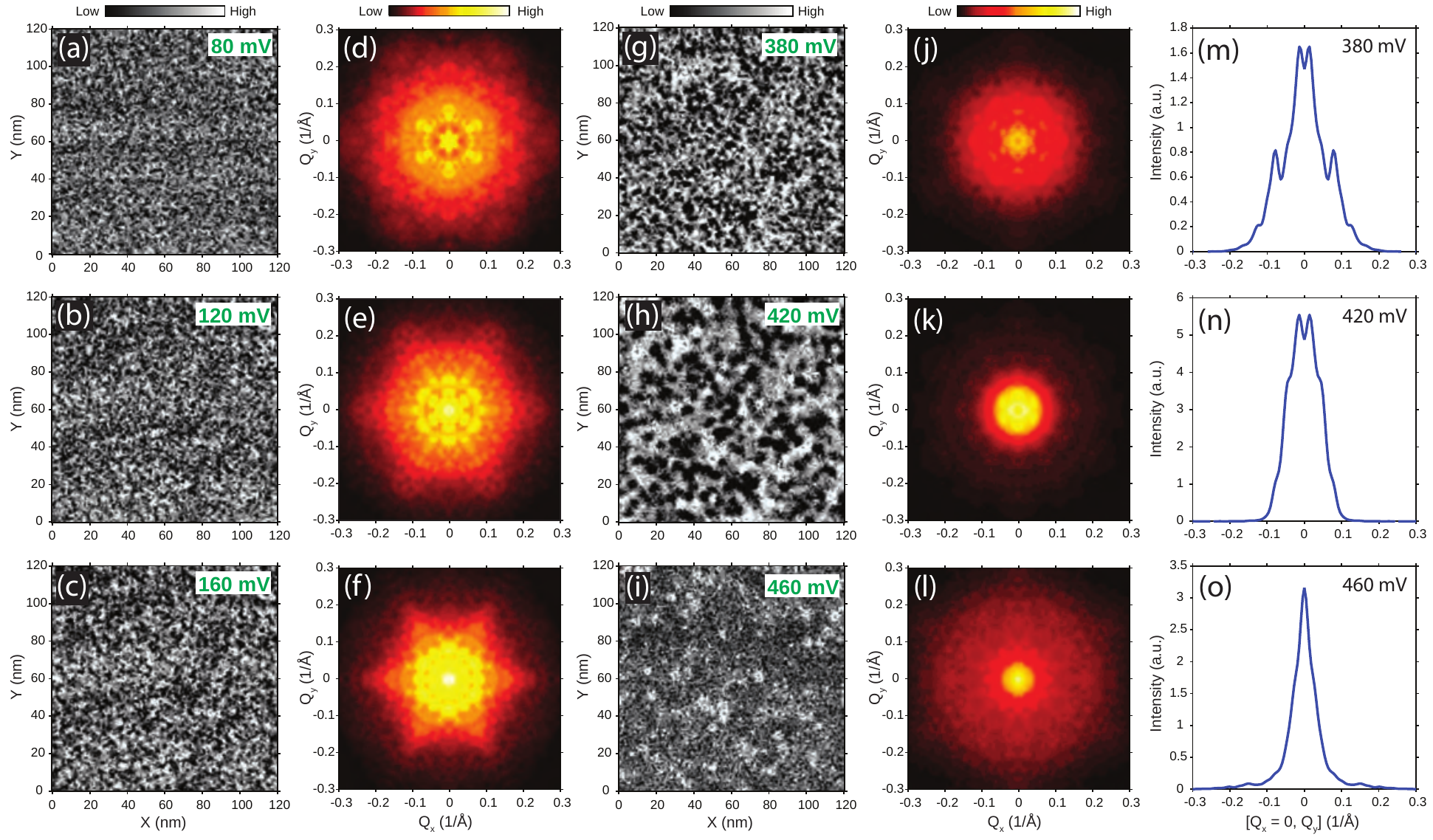}
\caption{\label{fig:3} (a-c) STS maps showing the real-space LDOS near the $170$\,meV Dirac point. (d-f) Fourier transforms of (a-c). (g-i) STS maps near the $420$\,meV Dirac point. (j-l) Fourier transforms of (g-i). (m-o) Line cuts along [$Q_x = 0$, $Q_y$]. The bias and current values were set to $800$\,mV and $500$\,pA respectively, and the $dI/dV$ signal was measured via a lock-in technique with a modulation amplitude of $10$\,mV. 
}
\end{figure*}

Figure \ref{fig:1}(a) shows the crystal structure of \hftep, where the atoms between consecutive Te layers along the c-axis are bonded by weak Van der Waals forces. This makes the Te layer the most likely termination resulting from the cleaving process~\cite{Hosen_2018}. 
Figure \ref{fig:1}(b) shows an atomically resolved STM constant-current topography on a cleaved surface of \hftep, featuring a hexagonal structure with lattice constants that are consistent with those obtained from x-ray diffraction~\cite{Chen_2016}. Localized defects are observed, appearing as depressions in the topography, as well as smoother features which may reflect underlying defects. We note that scanning multiple crystals, and many distinct microscopic regions within each cleave, we have only observed the Te-termination in the STM measurements. Using the same \textit{in situ} cleaving process, our ARPES measurements yield the occupied electronic structure of \hftep. 
Figure \ref{fig:1}(c) displays the result of the ARPES measured Fermi surface, which shows a hole-like pocket centered at the $\Gamma$ point and electron-like flower patterns, consistent with previous measurements. Overall, the momentum distribution curves (MDCs) show extremely sharp features which, together with the atomic resolution obtained in STM measurements, indicate the high quality of the surfaces obtained from the cleaving process \cite{SM_}. Figure\,\ref{fig:1}(d) shows a representative measurement of the differential conductance, $dI/dV$, which is proportional to the density of states as a function of energy. The most prominent feature in the spectrum is a pseudogap centered around $500$\,mV, with a small but non-zero density of states that could have surface and/or bulk origin. 

First we focus on SS2, which is predicted near $170$\,meV at the $\Gamma$ point~\cite{Hosen_2018}. To access states above $E_F$ with ARPES, we deposited K atoms on the cleaved surface of \hftep, which occupy the states above $E_F$ and enables their photo-emission~\cite{Hossain_2008}. 
Figures \ref{fig:2}(a-e) show the evolution of the constant energy contours for positive energies where zero refers to the Fermi energy of the sample before K-doping. The data in Fig.\,\ref{fig:2}(a-e) indicate the collapse of hole pocket into a single point at $\Gamma$, which is more clearly observed in the dispersion plot through $k_y = 0$, Fig.\,\ref{fig:2}(g), and could not be resolved without the K-doping -- compare to data in Fig.\,\ref{fig:2}(f), prior to K-doping. 

We also confirm the presence of the Dirac state at $170$\,meV via STS measurements of QPI patterns. Figures \ref{fig:3}(a-c) show $dI/dV$ maps for different energies above $E_F$ over a large field-of-view, $120$\,nm $\times$ $120$\,nm. 
It is evident from the STS maps that the characteristic length scales of the modulations in the local density of states (LDOS) are much larger than the inter-atomic distance and are strongly energy-dependent. 
This last behavior is typical of quasiparticle interference where the modulations are characterized by an energy dependent wave vector $\vec{Q}(E) = \vec{k}_f(E) - \vec{k}_i(E)$, where $\vec{k}_{i,f}(E)$ are determined by the band structure. The two-dimensional $Q$-space structure of the LDOS is obtained by Fourier transformation of the STS maps, $\tilde{g}(Q,E)$, as seen in Figs.\ref{fig:3}(d-f). Note that $\tilde{g}(Q,E=80$\,eV$)$ roughly features a hexagonal shape, while $\tilde{g}(Q,E=160$\,eV$)$ more closely matches a star shape, Fig.\,\ref{fig:3}(d-f), which more clearly conveys the energy dependence of the real-space QPI patterns.

\begin{figure}
\includegraphics[width=85mm]{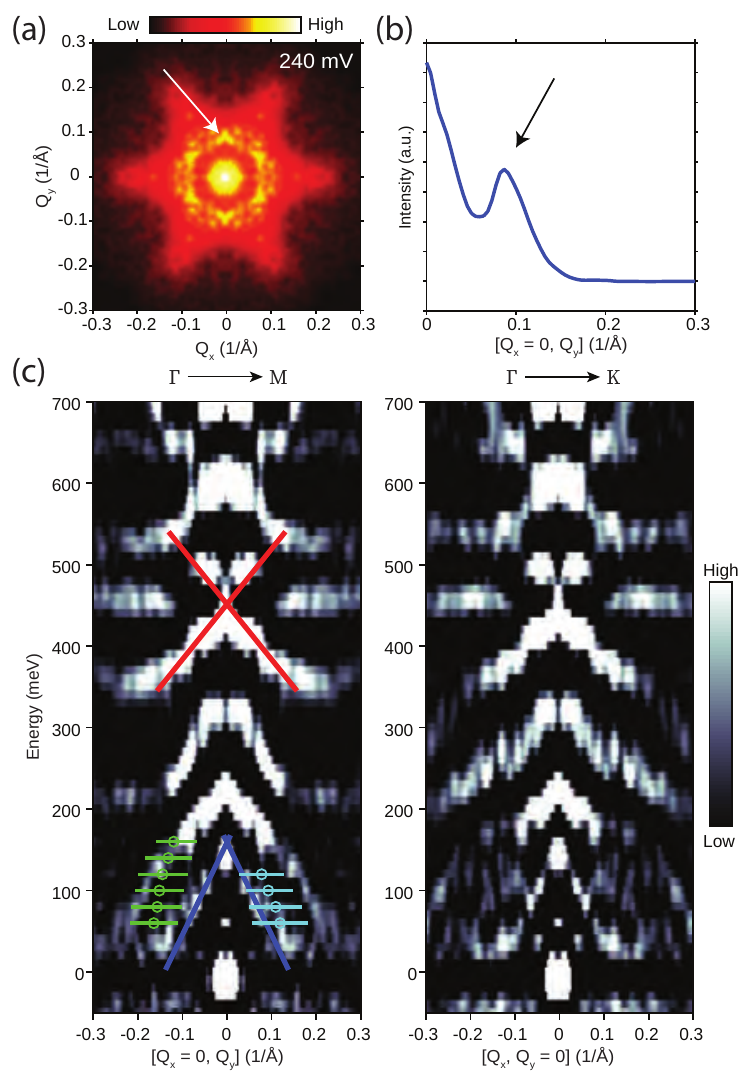}
\caption{\label{fig:4} (a) Fourier transform of the STS map at $240$\,mV. (b) Line cut of (a) along [$Q_x = 0$, $Q_y$], showing a distinct peak at $Q_y = 0.09$\,($1$/\AA). (c) QPI dispersion maps along the $\Gamma$-M and $\Gamma$-K directions. The ARPES measured Q-vectors are overlaid by cyan and green dots (with solid line error bars) for the dispersion curves displayed in Fig.\,\ref{fig:2}(h) and (j) respectively. Q-vector dispersions from the band-structure calculations along $\Gamma$-M are represented by the red and blue lines.
}
\end{figure}

We are able to quantitatively compare the results of the ARPES and STS measurements for energies between $E_F$ and $160$\,meV, which is the range accessed by both K-doped ARPES and STS. Line cuts of $\tilde{g}(Q,E)$ along the two high symmetry directions show distinct peak features on top of a large background. For example, a linecut along $[Q_x = 0$, $Q_y]$ at $240$\,meV reveals a distinct peak at $0.09$\,($1/$\AA), Fig.\,\ref{fig:4}(a-b). 
From $\tilde{g}(Q,E)$, we construct QPI dispersion maps along the two high symmetry directions, Fig.\,\ref{fig:4}(c), using the second-derivative method to clearly visualize the dispersive features. In the $0$ to $220$\,meV range we observe two peaks which disperse inwards as a function of energy. A detailed comparison to the ARPES data indicates that the inner dispersive peak matches the intra-pocket distance across $\Gamma$, while the outer peak dispersion more closely matches the separation between adjacent electron-pocket  \textit{petals} -- see the cyan and green circles in Fig.\,\ref{fig:4}(c), which correspond to the cyan and green lines in Figs.\,\ref{fig:2}(h,j). In fact, ARPES cuts along $[k_x, k_y=-0.53]$\,($1/$\AA) indicate a linear dispersion which extrapolate to a Dirac node near $220$\,meV, Figs.\,\ref{fig:2}(i,j). 
A similar state has been observed in \zrtep, with spin-orbit coupling expected to open a gap since the Dirac crossing is located away from time-reversal invariant momenta and is not protected by any crystalline symmetry \cite{ZrTeP_2016}. 
Our STS measurements do not resolve the presence of a gap at the $220$\,meV Dirac node, although our resolution was limited to $10$\,meV. [The $dI/dV$ was measured via a lock-in technique using a $10$\,mV modulation to the bias voltage.] Contrary to the outer peak, the inner QPI feature originates from the band centered at $\Gamma$, a time-reversal invariant point, as further confirmed by the excellent agreement with reported band structure calculations (blue line in Fig.\,\ref{fig:4}(c)) \cite{Hosen_2018}.    
Thus, our triple quantitative agreement between STS, ARPES and theory confirm the presence of a Dirac-like dispersion centered at $170$\,meV.

Finally, we focus on SS1, the strong topological SS predicted near $460$\,meV. Although ARPES cannot access this state, even with K-deposition, it can be resolved with STS measurements. First we note that although QPI signals can be the result of multiple surface or bulk bands, the calculated band structure of \hftep~indicates that the highest energy SS near $\Gamma$ exists inside a large bulk pseudogap ($\approx 0.13$\,eV), consistent with the observed minimum in the tunneling DOS, Fig.\,\ref{fig:1}(e). 
Focusing on that energy region, we observe a dramatic increase of the characteristic QPI wavelength as the bias is varied from $380$\,meV to $420$\,meV in our STS measurements, see Fig.\,\ref{fig:3}(g-h). The same trend is observed in the Fourier transforms, Fig.\,\ref{fig:3}(j-k) where a circular pattern shrinks towards $Q=0$, reflecting the expected QPI pattern from an isolated Dirac cone. Then, the expectation is that precisely at the Dirac point $\tilde{g}(Q)$ should display a single peak at $Q=0$, which becomes broad in the presence of disorder. Indeed, the real space STS data in Fig.\,\ref{fig:3}(i) shows a spatially inhomogeneous LDOS, 
due to the structural defects of the crystal, and a single central peak in the Fourier transform data, Fig.\,\ref{fig:3}(l). This trend is summarized in Figs.\,\ref{fig:3}(m-o), which display line cuts of the data in Figs.\,\ref{fig:3}(j-l), showing $Q_y \neq 0$ peaks converging to a single point at $460$\,meV. 
We note that in the absence of a predominant $Q\neq0$ QPI signal at the Dirac point, $\tilde{g}(Q,E=460$\,meV$)$ displays a weak halo, Fig.\,\ref{fig:3}(l,o), that stems from the intrinsic inhomogeneity of \hftep. Indeed, we find identical spatial patterns in both the $\tilde{g}(Q,E=460$\,meV$)$ and the topographic map over the same area \cite{SM_}, which confirms the non-QPI origin of the halo in Fig.\,\ref{fig:3}(l). Thus the intensity observed in the second-derivative map in Fig.\,\ref{fig:4}(c) at $E=460$\,meV and $Q\approx\pm0.17$\,($1/$\AA) is not indicative of an additional band. Now focusing on the energy-dependent QPI signal at lower $Q$ and comparing its energy dispersion to the prediction from calculations (red lines in Fig.\,\ref{fig:4}(c)), we find a remarkable agreement, thus demonstrating for the first time the presence of the Dirac surface state with a node at $460$\,meV above $E_F$. 

Our combined ARPES-STS measurements and their remarkable agreement with previous band structure calculations indicate the presence of TSSs above the Fermi level in \hftep. Although the topological nature of the various SSs and their spin-texture still requires more direct evidence from, for example,  spin-polarized ARPES~\cite{Hsieh_2010} or STS near extended defects~\cite{Seo2010}, the remarkable agreement to the reported band structure calculations strongly suggests their topological nature.
If these states are to be harnessed towards novel technologies they need to be tuned to $E_F$ by either chemical doping or surface gating. Here, the states above $E_F$ uncovered by our studies may provide a more direct path to these technologies than the previously measured states at $-0.9$\,eV. Also, the surface state centered at $460$\,meV is observed in the absence of other bands in our QPI measurement, suggesting a small contribution from the bulk, although only detailed measurements of the electronic bulk properties would confirm this.
Still, a high surface to bulk conductance ratio could be achieved in \hftep, since the strong topological nature of the Dirac surface state at $460$\,meV originates from the band inversion of Hf-\textit{d} and Te-\textit{p} orbitals, in contrast to traditional topological insulators. 
Note that the mixing Hf-\textit{d} and Te-\textit{p} orbital character, and its topology, may be tuned by chemical substitution, with the substitution of Hf (e.g. by Zr or Ti) controlling the \textit{d}-orbitals and the substitution of Te (e.g. by Se) regulating the \textit{p}-character -- not to mention the modification of lattice parameters by doping the P site (e.g. by As) or by chemical intercalations between layers.
Finally, since the current topological insulators  
suffer from contaminant conducting bulk channels that bypass the topologically protected surface states, the topological states uncovered by our studies may offer an alternative avenue for future applications.

\begin{acknowledgments}
This research used resources of the Advanced Light Source, which is a DOE Office of Science User Facility under contract no. DE-AC02-05CH11231. M. K. M. and V. T. acknowledge funding from the REU program NSF grant PHY-1560482.
\end{acknowledgments}

\bibliographystyle{apsrev4-1}
\bibliography{bib_lib}

\begin{thebibliography}{22}%
\makeatletter
\providecommand \@ifxundefined [1]{%
 \@ifx{#1\undefined}
}%
\providecommand \@ifnum [1]{%
 \ifnum #1\expandafter \@firstoftwo
 \else \expandafter \@secondoftwo
 \fi
}%
\providecommand \@ifx [1]{%
 \ifx #1\expandafter \@firstoftwo
 \else \expandafter \@secondoftwo
 \fi
}%
\providecommand \natexlab [1]{#1}%
\providecommand \enquote  [1]{``#1''}%
\providecommand \bibnamefont  [1]{#1}%
\providecommand \bibfnamefont [1]{#1}%
\providecommand \citenamefont [1]{#1}%
\providecommand \href@noop [0]{\@secondoftwo}%
\providecommand \href [0]{\begingroup \@sanitize@url \@href}%
\providecommand \@href[1]{\@@startlink{#1}\@@href}%
\providecommand \@@href[1]{\endgroup#1\@@endlink}%
\providecommand \@sanitize@url [0]{\catcode `\\12\catcode `\$12\catcode
  `\&12\catcode `\#12\catcode `\^12\catcode `\_12\catcode `\%12\relax}%
\providecommand \@@startlink[1]{}%
\providecommand \@@endlink[0]{}%
\providecommand \url  [0]{\begingroup\@sanitize@url \@url }%
\providecommand \@url [1]{\endgroup\@href {#1}{\urlprefix }}%
\providecommand \urlprefix  [0]{URL }%
\providecommand \Eprint [0]{\href }%
\providecommand \doibase [0]{http://dx.doi.org/}%
\providecommand \selectlanguage [0]{\@gobble}%
\providecommand \bibinfo  [0]{\@secondoftwo}%
\providecommand \bibfield  [0]{\@secondoftwo}%
\providecommand \translation [1]{[#1]}%
\providecommand \BibitemOpen [0]{}%
\providecommand \bibitemStop [0]{}%
\providecommand \bibitemNoStop [0]{.\EOS\space}%
\providecommand \EOS [0]{\spacefactor3000\relax}%
\providecommand \BibitemShut  [1]{\csname bibitem#1\endcsname}%
\let\auto@bib@innerbib\@empty
\bibitem [{\citenamefont {Qi}\ and\ \citenamefont {Zhang}(2011)}]{Qi_2011}%
  \BibitemOpen
  \bibfield  {author} {\bibinfo {author} {\bibfnamefont {X.-L.}\ \bibnamefont
  {Qi}}\ and\ \bibinfo {author} {\bibfnamefont {S.-C.}\ \bibnamefont {Zhang}},\
  }\href {\doibase 10.1103/RevModPhys.83.1057} {\bibfield  {journal} {\bibinfo
  {journal} {Rev. Mod. Phys.}\ }\textbf {\bibinfo {volume} {83}},\ \bibinfo
  {pages} {1057} (\bibinfo {year} {2011})}\BibitemShut {NoStop}%
\bibitem [{\citenamefont {Hasan}\ and\ \citenamefont
  {Kane}(2010)}]{Hasan_2010}%
  \BibitemOpen
  \bibfield  {author} {\bibinfo {author} {\bibfnamefont {M.~Z.}\ \bibnamefont
  {Hasan}}\ and\ \bibinfo {author} {\bibfnamefont {C.~L.}\ \bibnamefont
  {Kane}},\ }\href {\doibase 10.1103/RevModPhys.82.3045} {\bibfield  {journal}
  {\bibinfo  {journal} {Rev. Mod. Phys.}\ }\textbf {\bibinfo {volume} {82}},\
  \bibinfo {pages} {3045} (\bibinfo {year} {2010})}\BibitemShut {NoStop}%
\bibitem [{\citenamefont {Ando}(2013)}]{Ando_2013}%
  \BibitemOpen
  \bibfield  {author} {\bibinfo {author} {\bibfnamefont {Y.}~\bibnamefont
  {Ando}},\ }\href {\doibase 10.7566/JPSJ.82.102001} {\bibfield  {journal}
  {\bibinfo  {journal} {Journal of the Physical Society of Japan}\ }\textbf
  {\bibinfo {volume} {82}},\ \bibinfo {pages} {102001} (\bibinfo {year}
  {2013})}\BibitemShut {NoStop}%
\bibitem [{\citenamefont {Armitage}\ \emph {et~al.}(2018)\citenamefont
  {Armitage}, \citenamefont {Mele},\ and\ \citenamefont
  {Vishwanath}}]{Armitage_2018}%
  \BibitemOpen
  \bibfield  {author} {\bibinfo {author} {\bibfnamefont {N.~P.}\ \bibnamefont
  {Armitage}}, \bibinfo {author} {\bibfnamefont {E.~J.}\ \bibnamefont {Mele}},
  \ and\ \bibinfo {author} {\bibfnamefont {A.}~\bibnamefont {Vishwanath}},\
  }\href {\doibase 10.1103/RevModPhys.90.015001} {\bibfield  {journal}
  {\bibinfo  {journal} {Rev. Mod. Phys.}\ }\textbf {\bibinfo {volume} {90}},\
  \bibinfo {pages} {015001} (\bibinfo {year} {2018})}\BibitemShut {NoStop}%
\bibitem [{\citenamefont {Yang}\ \emph {et~al.}(2018)\citenamefont {Yang},
  \citenamefont {Yang}, \citenamefont {Derunova}, \citenamefont {Parkin},
  \citenamefont {Yan},\ and\ \citenamefont {Ali}}]{Yang_2018}%
  \BibitemOpen
  \bibfield  {author} {\bibinfo {author} {\bibfnamefont {S.-Y.}\ \bibnamefont
  {Yang}}, \bibinfo {author} {\bibfnamefont {H.}~\bibnamefont {Yang}}, \bibinfo
  {author} {\bibfnamefont {E.}~\bibnamefont {Derunova}}, \bibinfo {author}
  {\bibfnamefont {S.~S.~P.}\ \bibnamefont {Parkin}}, \bibinfo {author}
  {\bibfnamefont {B.}~\bibnamefont {Yan}}, \ and\ \bibinfo {author}
  {\bibfnamefont {M.~N.}\ \bibnamefont {Ali}},\ }\href {\doibase
  10.1080/23746149.2017.1414631} {\bibfield  {journal} {\bibinfo  {journal}
  {Advances in Physics: X}\ }\textbf {\bibinfo {volume} {3}},\ \bibinfo {pages}
  {1414631} (\bibinfo {year} {2018})}\BibitemShut {NoStop}%
\bibitem [{\citenamefont {Alicea}(2012)}]{Alicea_2012}%
  \BibitemOpen
  \bibfield  {author} {\bibinfo {author} {\bibfnamefont {J.}~\bibnamefont
  {Alicea}},\ }\href {\doibase 10.1088/0034-4885/75/7/076501} {\bibfield
  {journal} {\bibinfo  {journal} {Reports on Progress in Physics}\ }\textbf
  {\bibinfo {volume} {75}},\ \bibinfo {pages} {076501} (\bibinfo {year}
  {2012})}\BibitemShut {NoStop}%
\bibitem [{\citenamefont {{\v{S}}mejkal}\ \emph {et~al.}(2018)\citenamefont
  {{\v{S}}mejkal}, \citenamefont {Mokrousov}, \citenamefont {Yan},\ and\
  \citenamefont {MacDonald}}]{Smejkal_2018}%
  \BibitemOpen
  \bibfield  {author} {\bibinfo {author} {\bibfnamefont {L.}~\bibnamefont
  {{\v{S}}mejkal}}, \bibinfo {author} {\bibfnamefont {Y.}~\bibnamefont
  {Mokrousov}}, \bibinfo {author} {\bibfnamefont {B.}~\bibnamefont {Yan}}, \
  and\ \bibinfo {author} {\bibfnamefont {A.~H.}\ \bibnamefont {MacDonald}},\
  }\href {\doibase 10.1038/s41567-018-0064-5} {\bibfield  {journal} {\bibinfo
  {journal} {Nature Physics}\ }\textbf {\bibinfo {volume} {14}},\ \bibinfo
  {pages} {242} (\bibinfo {year} {2018})}\BibitemShut {NoStop}%
\bibitem [{\citenamefont {Ji}\ \emph {et~al.}(2016)\citenamefont {Ji},
  \citenamefont {Pletikosi\ifmmode~\acute{c}\else \'{c}\fi{}}, \citenamefont
  {Gibson}, \citenamefont {Sahasrabudhe}, \citenamefont {Valla},\ and\
  \citenamefont {Cava}}]{ZrTeP_2016}%
  \BibitemOpen
  \bibfield  {author} {\bibinfo {author} {\bibfnamefont {H.}~\bibnamefont
  {Ji}}, \bibinfo {author} {\bibfnamefont {I.}~\bibnamefont
  {Pletikosi\ifmmode~\acute{c}\else \'{c}\fi{}}}, \bibinfo {author}
  {\bibfnamefont {Q.~D.}\ \bibnamefont {Gibson}}, \bibinfo {author}
  {\bibfnamefont {G.}~\bibnamefont {Sahasrabudhe}}, \bibinfo {author}
  {\bibfnamefont {T.}~\bibnamefont {Valla}}, \ and\ \bibinfo {author}
  {\bibfnamefont {R.~J.}\ \bibnamefont {Cava}},\ }\href {\doibase
  10.1103/PhysRevB.93.045315} {\bibfield  {journal} {\bibinfo  {journal} {Phys.
  Rev. B}\ }\textbf {\bibinfo {volume} {93}},\ \bibinfo {pages} {045315}
  (\bibinfo {year} {2016})}\BibitemShut {NoStop}%
\bibitem [{\citenamefont {Hosen}\ \emph {et~al.}(2018)\citenamefont {Hosen},
  \citenamefont {Dimitri}, \citenamefont {Kumar~Nandy}, \citenamefont {Aperis},
  \citenamefont {Sankar}, \citenamefont {Dhakal}, \citenamefont {Maldonado},
  \citenamefont {Kabir}, \citenamefont {Sims}, \citenamefont {Chou},
  \citenamefont {Kaczorowski}, \citenamefont {Durakiewicz}, \citenamefont
  {Oppeneer},\ and\ \citenamefont {Neupane}}]{Hosen_2018}%
  \BibitemOpen
  \bibfield  {author} {\bibinfo {author} {\bibfnamefont {M.~M.}\ \bibnamefont
  {Hosen}}, \bibinfo {author} {\bibfnamefont {K.}~\bibnamefont {Dimitri}},
  \bibinfo {author} {\bibfnamefont {A.}~\bibnamefont {Kumar~Nandy}}, \bibinfo
  {author} {\bibfnamefont {A.}~\bibnamefont {Aperis}}, \bibinfo {author}
  {\bibfnamefont {R.}~\bibnamefont {Sankar}}, \bibinfo {author} {\bibfnamefont
  {G.}~\bibnamefont {Dhakal}}, \bibinfo {author} {\bibfnamefont
  {P.}~\bibnamefont {Maldonado}}, \bibinfo {author} {\bibfnamefont
  {F.}~\bibnamefont {Kabir}}, \bibinfo {author} {\bibfnamefont
  {C.}~\bibnamefont {Sims}}, \bibinfo {author} {\bibfnamefont {F.}~\bibnamefont
  {Chou}}, \bibinfo {author} {\bibfnamefont {D.}~\bibnamefont {Kaczorowski}},
  \bibinfo {author} {\bibfnamefont {T.}~\bibnamefont {Durakiewicz}}, \bibinfo
  {author} {\bibfnamefont {P.}~\bibnamefont {Oppeneer}}, \ and\ \bibinfo
  {author} {\bibfnamefont {M.}~\bibnamefont {Neupane}},\ }\href {\doibase
  10.1038/s41467-018-05233-1} {\bibfield  {journal} {\bibinfo  {journal}
  {Nature Communications}\ }\textbf {\bibinfo {volume} {9}},\ \bibinfo {pages}
  {3002} (\bibinfo {year} {2018})}\BibitemShut {NoStop}%
\bibitem [{\citenamefont {Fu}\ and\ \citenamefont {Kane}(2007)}]{Fu_2007}%
  \BibitemOpen
  \bibfield  {author} {\bibinfo {author} {\bibfnamefont {L.}~\bibnamefont
  {Fu}}\ and\ \bibinfo {author} {\bibfnamefont {C.~L.}\ \bibnamefont {Kane}},\
  }\href {\doibase 10.1103/PhysRevB.76.045302} {\bibfield  {journal} {\bibinfo
  {journal} {Phys. Rev. B}\ }\textbf {\bibinfo {volume} {76}},\ \bibinfo
  {pages} {045302} (\bibinfo {year} {2007})}\BibitemShut {NoStop}%
\bibitem [{\citenamefont {Yang}\ and\ \citenamefont {Liu}(2014)}]{Yang2014}%
  \BibitemOpen
  \bibfield  {author} {\bibinfo {author} {\bibfnamefont {M.}~\bibnamefont
  {Yang}}\ and\ \bibinfo {author} {\bibfnamefont {W.-M.}\ \bibnamefont {Liu}},\
  }\href {https://doi.org/10.1038/srep05131} {\bibfield  {journal} {\bibinfo
  {journal} {Scientific Reports}\ }\textbf {\bibinfo {volume} {4}},\ \bibinfo
  {pages} {5131} (\bibinfo {year} {2014})}\BibitemShut {NoStop}%
\bibitem [{\citenamefont {Hoffman}\ \emph {et~al.}(2002)\citenamefont
  {Hoffman}, \citenamefont {McElroy}, \citenamefont {Lee}, \citenamefont
  {Lang}, \citenamefont {Eisaki}, \citenamefont {Uchida},\ and\ \citenamefont
  {Davis}}]{Hoffman_QPI_2002}%
  \BibitemOpen
  \bibfield  {author} {\bibinfo {author} {\bibfnamefont {J.~E.}\ \bibnamefont
  {Hoffman}}, \bibinfo {author} {\bibfnamefont {K.}~\bibnamefont {McElroy}},
  \bibinfo {author} {\bibfnamefont {D.-H.}\ \bibnamefont {Lee}}, \bibinfo
  {author} {\bibfnamefont {K.~M.}\ \bibnamefont {Lang}}, \bibinfo {author}
  {\bibfnamefont {H.}~\bibnamefont {Eisaki}}, \bibinfo {author} {\bibfnamefont
  {S.}~\bibnamefont {Uchida}}, \ and\ \bibinfo {author} {\bibfnamefont {J.~C.}\
  \bibnamefont {Davis}},\ }\href {\doibase 10.1126/science.1072640} {\bibfield
  {journal} {\bibinfo  {journal} {Science}\ }\textbf {\bibinfo {volume}
  {297}},\ \bibinfo {pages} {1148} (\bibinfo {year} {2002})}\BibitemShut
  {NoStop}%
\bibitem [{\citenamefont {Wang}\ and\ \citenamefont
  {Lee}(2003)}]{Wang_Lee_QPI}%
  \BibitemOpen
  \bibfield  {author} {\bibinfo {author} {\bibfnamefont {Q.-H.}\ \bibnamefont
  {Wang}}\ and\ \bibinfo {author} {\bibfnamefont {D.-H.}\ \bibnamefont {Lee}},\
  }\href {\doibase 10.1103/PhysRevB.67.020511} {\bibfield  {journal} {\bibinfo
  {journal} {Phys. Rev. B}\ }\textbf {\bibinfo {volume} {67}},\ \bibinfo
  {pages} {020511(R)} (\bibinfo {year} {2003})}\BibitemShut {NoStop}%
\bibitem [{\citenamefont {Roushan}\ \emph {et~al.}(2009)\citenamefont
  {Roushan}, \citenamefont {Seo}, \citenamefont {Parker}, \citenamefont {Hor},
  \citenamefont {Hsieh}, \citenamefont {Qian}, \citenamefont {Richardella},
  \citenamefont {Hasan}, \citenamefont {Cava},\ and\ \citenamefont
  {Yazdani}}]{Roushan_Nature_2009}%
  \BibitemOpen
  \bibfield  {author} {\bibinfo {author} {\bibfnamefont {P.}~\bibnamefont
  {Roushan}}, \bibinfo {author} {\bibfnamefont {J.}~\bibnamefont {Seo}},
  \bibinfo {author} {\bibfnamefont {C.~V.}\ \bibnamefont {Parker}}, \bibinfo
  {author} {\bibfnamefont {Y.~S.}\ \bibnamefont {Hor}}, \bibinfo {author}
  {\bibfnamefont {D.}~\bibnamefont {Hsieh}}, \bibinfo {author} {\bibfnamefont
  {D.}~\bibnamefont {Qian}}, \bibinfo {author} {\bibfnamefont {A.}~\bibnamefont
  {Richardella}}, \bibinfo {author} {\bibfnamefont {M.~Z.}\ \bibnamefont
  {Hasan}}, \bibinfo {author} {\bibfnamefont {R.~J.}\ \bibnamefont {Cava}}, \
  and\ \bibinfo {author} {\bibfnamefont {A.}~\bibnamefont {Yazdani}},\ }\href
  {https://doi.org/10.1038/nature08308} {\bibfield  {journal} {\bibinfo
  {journal} {Nature}\ }\textbf {\bibinfo {volume} {460}},\ \bibinfo {pages}
  {1106} (\bibinfo {year} {2009})}\BibitemShut {NoStop}%
\bibitem [{\citenamefont {Zeljkovic}\ \emph {et~al.}(2014)\citenamefont
  {Zeljkovic}, \citenamefont {Okada}, \citenamefont {Huang}, \citenamefont
  {Sankar}, \citenamefont {Walkup}, \citenamefont {Zhou}, \citenamefont
  {Serbyn}, \citenamefont {Chou}, \citenamefont {Tsai}, \citenamefont {Lin},
  \citenamefont {Bansil}, \citenamefont {Fu}, \citenamefont {Hasan},\ and\
  \citenamefont {Madhavan}}]{Zeljkovic2014}%
  \BibitemOpen
  \bibfield  {author} {\bibinfo {author} {\bibfnamefont {I.}~\bibnamefont
  {Zeljkovic}}, \bibinfo {author} {\bibfnamefont {Y.}~\bibnamefont {Okada}},
  \bibinfo {author} {\bibfnamefont {C.-Y.}\ \bibnamefont {Huang}}, \bibinfo
  {author} {\bibfnamefont {R.}~\bibnamefont {Sankar}}, \bibinfo {author}
  {\bibfnamefont {D.}~\bibnamefont {Walkup}}, \bibinfo {author} {\bibfnamefont
  {W.}~\bibnamefont {Zhou}}, \bibinfo {author} {\bibfnamefont {M.}~\bibnamefont
  {Serbyn}}, \bibinfo {author} {\bibfnamefont {F.}~\bibnamefont {Chou}},
  \bibinfo {author} {\bibfnamefont {W.-F.}\ \bibnamefont {Tsai}}, \bibinfo
  {author} {\bibfnamefont {H.}~\bibnamefont {Lin}}, \bibinfo {author}
  {\bibfnamefont {A.}~\bibnamefont {Bansil}}, \bibinfo {author} {\bibfnamefont
  {L.}~\bibnamefont {Fu}}, \bibinfo {author} {\bibfnamefont {M.~Z.}\
  \bibnamefont {Hasan}}, \ and\ \bibinfo {author} {\bibfnamefont
  {V.}~\bibnamefont {Madhavan}},\ }\href {https://doi.org/10.1038/nphys3012}
  {\bibfield  {journal} {\bibinfo  {journal} {Nature Physics}\ }\textbf
  {\bibinfo {volume} {10}},\ \bibinfo {pages} {572} (\bibinfo {year}
  {2014})}\BibitemShut {NoStop}%
\bibitem [{\citenamefont {Yazdani}\ \emph {et~al.}(2016)\citenamefont
  {Yazdani}, \citenamefont {da~Silva~Neto},\ and\ \citenamefont
  {Aynajian}}]{Yazdani_review}%
  \BibitemOpen
  \bibfield  {author} {\bibinfo {author} {\bibfnamefont {A.}~\bibnamefont
  {Yazdani}}, \bibinfo {author} {\bibfnamefont {E.~H.}\ \bibnamefont
  {da~Silva~Neto}}, \ and\ \bibinfo {author} {\bibfnamefont {P.}~\bibnamefont
  {Aynajian}},\ }\href {\doibase 10.1146/annurev-conmatphys-031214-014529}
  {\bibfield  {journal} {\bibinfo  {journal} {Annual Review of Condensed Matter
  Physics}\ }\textbf {\bibinfo {volume} {7}},\ \bibinfo {pages} {11} (\bibinfo
  {year} {2016})},\ \Eprint
  {http://arxiv.org/abs/https://doi.org/10.1146/annurev-conmatphys-031214-014529}
  {https://doi.org/10.1146/annurev-conmatphys-031214-014529} \BibitemShut
  {NoStop}%
\bibitem [{\citenamefont {Tschulik}\ \emph {et~al.}(2009)\citenamefont
  {Tschulik}, \citenamefont {Ruck}, \citenamefont {Binnewies}, \citenamefont
  {Milke}, \citenamefont {Hoffmann}, \citenamefont {Schnelle}, \citenamefont
  {Fokwa}, \citenamefont {Gilleßen},\ and\ \citenamefont
  {Schmidt}}]{Tschulik_2009}%
  \BibitemOpen
  \bibfield  {author} {\bibinfo {author} {\bibfnamefont {K.}~\bibnamefont
  {Tschulik}}, \bibinfo {author} {\bibfnamefont {M.}~\bibnamefont {Ruck}},
  \bibinfo {author} {\bibfnamefont {M.}~\bibnamefont {Binnewies}}, \bibinfo
  {author} {\bibfnamefont {E.}~\bibnamefont {Milke}}, \bibinfo {author}
  {\bibfnamefont {S.}~\bibnamefont {Hoffmann}}, \bibinfo {author}
  {\bibfnamefont {W.}~\bibnamefont {Schnelle}}, \bibinfo {author}
  {\bibfnamefont {B.~P.~T.}\ \bibnamefont {Fokwa}}, \bibinfo {author}
  {\bibfnamefont {M.}~\bibnamefont {Gilleßen}}, \ and\ \bibinfo {author}
  {\bibfnamefont {P.}~\bibnamefont {Schmidt}},\ }\href {\doibase
  10.1002/ejic.200900346} {\bibfield  {journal} {\bibinfo  {journal} {European
  Journal of Inorganic Chemistry}\ }\textbf {\bibinfo {volume} {2009}},\
  \bibinfo {pages} {3102} (\bibinfo {year} {2009})}\BibitemShut {NoStop}%
\bibitem [{SM_()}]{SM_}%
  \BibitemOpen
  \href@noop {} {\bibinfo  {journal} {See Supplementary Materials}\
  }\BibitemShut {NoStop}%
\bibitem [{\citenamefont {Chen}\ \emph {et~al.}(2016)\citenamefont {Chen},
  \citenamefont {Das}, \citenamefont {Rhodes}, \citenamefont {Memaran},
  \citenamefont {Besara}, \citenamefont {Siegrist}, \citenamefont {Manousakis},
  \citenamefont {Balicas},\ and\ \citenamefont {Baumbach}}]{Chen_2016}%
  \BibitemOpen
\bibfield  {journal} {  }\bibfield  {author} {\bibinfo {author} {\bibfnamefont
  {K.-W.}\ \bibnamefont {Chen}}, \bibinfo {author} {\bibfnamefont
  {S.}~\bibnamefont {Das}}, \bibinfo {author} {\bibfnamefont {D.}~\bibnamefont
  {Rhodes}}, \bibinfo {author} {\bibfnamefont {S.}~\bibnamefont {Memaran}},
  \bibinfo {author} {\bibfnamefont {T.}~\bibnamefont {Besara}}, \bibinfo
  {author} {\bibfnamefont {T.}~\bibnamefont {Siegrist}}, \bibinfo {author}
  {\bibfnamefont {E.}~\bibnamefont {Manousakis}}, \bibinfo {author}
  {\bibfnamefont {L.}~\bibnamefont {Balicas}}, \ and\ \bibinfo {author}
  {\bibfnamefont {R.~E.}\ \bibnamefont {Baumbach}},\ }\href {\doibase
  10.1088/0953-8984/28/14/14lt01} {\bibfield  {journal} {\bibinfo  {journal}
  {Journal of Physics: Condensed Matter}\ }\textbf {\bibinfo {volume} {28}},\
  \bibinfo {pages} {14LT01} (\bibinfo {year} {2016})}\BibitemShut {NoStop}%
\bibitem [{\citenamefont {Hossain}\ \emph {et~al.}(2008)\citenamefont
  {Hossain}, \citenamefont {Mottershead}, \citenamefont {Fournier},
  \citenamefont {Bostwick}, \citenamefont {McChesney}, \citenamefont
  {Rotenberg}, \citenamefont {Liang}, \citenamefont {Hardy}, \citenamefont
  {Sawatzky}, \citenamefont {Elfimov}, \citenamefont {Bonn},\ and\
  \citenamefont {Damascelli}}]{Hossain_2008}%
  \BibitemOpen
  \bibfield  {author} {\bibinfo {author} {\bibfnamefont {M.~A.}\ \bibnamefont
  {Hossain}}, \bibinfo {author} {\bibfnamefont {J.~D.~F.}\ \bibnamefont
  {Mottershead}}, \bibinfo {author} {\bibfnamefont {D.}~\bibnamefont
  {Fournier}}, \bibinfo {author} {\bibfnamefont {A.}~\bibnamefont {Bostwick}},
  \bibinfo {author} {\bibfnamefont {J.~L.}\ \bibnamefont {McChesney}}, \bibinfo
  {author} {\bibfnamefont {E.}~\bibnamefont {Rotenberg}}, \bibinfo {author}
  {\bibfnamefont {R.}~\bibnamefont {Liang}}, \bibinfo {author} {\bibfnamefont
  {W.~N.}\ \bibnamefont {Hardy}}, \bibinfo {author} {\bibfnamefont {G.~A.}\
  \bibnamefont {Sawatzky}}, \bibinfo {author} {\bibfnamefont {I.~S.}\
  \bibnamefont {Elfimov}}, \bibinfo {author} {\bibfnamefont {D.~A.}\
  \bibnamefont {Bonn}}, \ and\ \bibinfo {author} {\bibfnamefont
  {A.}~\bibnamefont {Damascelli}},\ }\href {https://doi.org/10.1038/nphys998
  http://10.0.4.14/nphys998
  https://www.nature.com/articles/nphys998{\#}supplementary-information}
  {\bibfield  {journal} {\bibinfo  {journal} {Nature Physics}\ }\textbf
  {\bibinfo {volume} {4}},\ \bibinfo {pages} {527} (\bibinfo {year}
  {2008})}\BibitemShut {NoStop}%
\bibitem [{\citenamefont {Hsieh}\ \emph {et~al.}(2010)\citenamefont {Hsieh},
  \citenamefont {Wray}, \citenamefont {Qian}, \citenamefont {Xia},
  \citenamefont {Dil}, \citenamefont {Meier}, \citenamefont {Patthey},
  \citenamefont {Osterwalder}, \citenamefont {Bihlmayer}, \citenamefont {Hor},
  \citenamefont {Cava},\ and\ \citenamefont {Hasan}}]{Hsieh_2010}%
  \BibitemOpen
  \bibfield  {author} {\bibinfo {author} {\bibfnamefont {D.}~\bibnamefont
  {Hsieh}}, \bibinfo {author} {\bibfnamefont {L.}~\bibnamefont {Wray}},
  \bibinfo {author} {\bibfnamefont {D.}~\bibnamefont {Qian}}, \bibinfo {author}
  {\bibfnamefont {Y.}~\bibnamefont {Xia}}, \bibinfo {author} {\bibfnamefont
  {J.~H.}\ \bibnamefont {Dil}}, \bibinfo {author} {\bibfnamefont
  {F.}~\bibnamefont {Meier}}, \bibinfo {author} {\bibfnamefont
  {L.}~\bibnamefont {Patthey}}, \bibinfo {author} {\bibfnamefont
  {J.}~\bibnamefont {Osterwalder}}, \bibinfo {author} {\bibfnamefont
  {G.}~\bibnamefont {Bihlmayer}}, \bibinfo {author} {\bibfnamefont {Y.~S.}\
  \bibnamefont {Hor}}, \bibinfo {author} {\bibfnamefont {R.~J.}\ \bibnamefont
  {Cava}}, \ and\ \bibinfo {author} {\bibfnamefont {M.~Z.}\ \bibnamefont
  {Hasan}},\ }\href {\doibase 10.1088/1367-2630/12/12/125001} {\bibfield
  {journal} {\bibinfo  {journal} {New Journal of Physics}\ }\textbf {\bibinfo
  {volume} {12}},\ \bibinfo {pages} {125001} (\bibinfo {year}
  {2010})}\BibitemShut {NoStop}%
\bibitem [{\citenamefont {Seo}\ \emph {et~al.}(2010)\citenamefont {Seo},
  \citenamefont {Roushan}, \citenamefont {Beidenkopf}, \citenamefont {Hor},
  \citenamefont {Cava},\ and\ \citenamefont {Yazdani}}]{Seo2010}%
  \BibitemOpen
  \bibfield  {author} {\bibinfo {author} {\bibfnamefont {J.}~\bibnamefont
  {Seo}}, \bibinfo {author} {\bibfnamefont {P.}~\bibnamefont {Roushan}},
  \bibinfo {author} {\bibfnamefont {H.}~\bibnamefont {Beidenkopf}}, \bibinfo
  {author} {\bibfnamefont {Y.~S.}\ \bibnamefont {Hor}}, \bibinfo {author}
  {\bibfnamefont {R.~J.}\ \bibnamefont {Cava}}, \ and\ \bibinfo {author}
  {\bibfnamefont {A.}~\bibnamefont {Yazdani}},\ }\href
  {https://doi.org/10.1038/nature09189} {\bibfield  {journal} {\bibinfo
  {journal} {Nature}\ }\textbf {\bibinfo {volume} {466}},\ \bibinfo {pages}
  {343} (\bibinfo {year} {2010})}\BibitemShut {NoStop}%
\end{thebibliography}%
\end{document}